\documentclass[10pt,fleqn]{article}

\usepackage{amsmath,amssymb}
\usepackage{graphicx}
\usepackage[top=0.75in, bottom=0.75in, left=0.75in, right=0.75in, dvips]{geometry}

\begin{document}

\title{Radiative Electroweak Symmetry Breaking Beyond Leading Logarithms}

\author{V.\ Elias\thanks{Department of Applied Mathematics, The University of Western Ontario, London, Ontario  N6A 5B7, Canada},
R.B.\ Mann\thanks{
Department of Physics, University of Waterloo,
Waterloo, Ontario  N2L 3G1,  Canada}, 
D.G.C.\ McKeon\thanks{Department of Applied Mathematics, The University of Western Ontario, London, Ontario  N6A 5B7, Canada},
T.G.\ Steele\thanks{Department of Physics and Engineering Physics, University of
Saskatchewan, Saskatoon, SK, S7N 5E2, Canada}
}
\maketitle
\begin{abstract}
The top-quark Yukawa coupling is too large to permit radiative electroweak symmetry breaking to occur for small values of $y$, the Higgs self-coupling, to leading-logarithm order.  However,  a large $y$ solution leading to a viable Higgs mass of approximately $220\,{\rm GeV}$ does exist, and differs from conventional symmetry breaking by an approximately five-fold enhancement of the Higgs self-coupling.  This scenario for radiative symmetry breaking is reviewed, and the order-by-order perturbative stability of this scenario is studied within the scalar field theory projection of the standard model in which  the Higgs self-coupling $y$ represents the dominant standard-model coupling.
\end{abstract}
%

Conventional electroweak (EW) symmetry breaking requires the presence of a Higgs scalar-field quadratic term in the Lagrangian.  Such a mass term is unnatural if $SU(2)\times U(1)$ EW gauge theory is embedded
within a grand-unified theory (GUT), since fine-tuning is needed to cancel the unification-scale perturbative corrections generated by this mass term \cite{sher} to maintain a Higgs mass empirically bounded not too far from the EW vacuum expectation value (VEV) scale $\langle \phi\rangle=v=246.2\,{\rm GeV}$ \cite{higgs_mass}.  This fine-tuning problem can be circumvented if the embedding unified theory has a symmetry ({\it e.g.} conformal symmetry) which protects these quadratic terms from GUT-scale corrections. 
Radiative EW symmetry breaking provides such a scenario. Quadratic scalar mass terms are absent, and in the seminal work of Coleman \& Weinberg, it is demonstrated that spontaneous symmetry breaking ({\it i.e.}
the generation of a VEV) occurs via radiative (perturbative) corrections to the conformally-invariant theory \cite{CW}. 

Unlike conventional symmetry breaking where the Higgs mass is an unconstrained parameter, the radiative symmetry breaking mechanism is a self-consistent approach which results in a prediction of both the Higgs mass and its four-point self-coupling after imposition of the external EW scale for the VEV $\langle \phi\rangle=v=246.2\,{GeV}$.
In the absence of large Yukawa couplings ({\it i.e.} Yukawa couplings are dominated by EW gauge couplings), a justifiable assumption at the time of Coleman \& Weinberg's work, the radiative symmetry breaking scenario contains a small-$\lambda$ solution leading to an ${\cal O}(10\,{\rm GeV})$ Higgs mass \cite{CW}, long since ruled out via direct experimental searches.  However, the top quark Yukawa coupling is large enough to destabilize this small-$\lambda$ solution.  
The Coleman-Weinberg radiative mechanism has thus been revisited in the context of the large top-quark Yukawa coupling, revealing the persistence of a large-$\lambda$ radiative scenario resulting in a $218\,{\rm GeV}$ Higgs mass for the minimal (single-Higgs-doublet) standard model \cite{us1,us2}.  

Consider the one-loop effective potential $V_{eff}=\pi^2\phi^4 S$ for the Higgs sector, which must satisfy the renormalization group (RG) equation
\begin{equation}
0=\mu\frac{dV_{eff}}{d\mu}=\left[ \left( -2-2\gamma \right) \frac{\partial }{\partial L}+\beta _{y}%
\frac{\partial }{\partial y}+\sum_{k=\{x,z,r,s\}}\beta _{k}\frac{\partial }{\partial k}-4\gamma \right] S=0~,  
\label{rg_eq}
\end{equation}%
where $L=\log\left(\phi^2/\mu^2\right)$, $x=g_t^2/4\pi^2$ represents the dominant Yukawa coupling,  $\{z=g_3^2/4\pi^2,r=g_2^2/4\pi^2,s={g'}^2/4\pi^2\}$
respectively represent the $\{SU(3),SU(2),U(1)\}$ standard model gauge couplings, and  the quantity $y=\lambda/4\pi^2$   is referenced to the tree-level potential $V=\lambda \phi^4/4$.  Given the expressions \cite{sher,rg_ref} for the RG functions $\beta_k$ and $\gamma$,  the one-loop solution to (\ref{rg_eq}) is
\begin{equation}
V_{eff}^{1L}(\phi)=\pi^2\phi^4\left[y+\Omega L\right]+ \pi^2\phi^4 K~,~\Omega=3y^2-\frac{3}{4}x^2+\frac{9r^2+6rs+3s^2}{64}~,
\label{V_oneloop}
\end{equation}
where $K$ (combined with the $\phi^4$ factor) represents a logarithm-independent  counter-term.  Choosing the renormalization scale $\mu=v$, requiring $V_{eff}^{\prime }(v)=0$ so that the effective potential has a minimum at $\phi=v$ , and imposing the renormalization condition $V_{eff}^{(4)}(v)=V_{tree}^{(4)}(v)=24\pi^2y$ \cite{CW} results in a constraint equation among the various couplings:
\begin{equation}
y=\frac{11}{3}\left[3y^2+\frac{3}{64}\left(s^2+2rs+3r^2\right)-\frac{3}{4}x^2 \right] ~.
\label{lo_const}
\end{equation}
Given the physical values of the couplings $\{z(v)=0.0329,~r(v)=0.0109,~s(v)=0.00324\}$ at the EW scale $v$, it is easy to see that only the top-quark Yukawa coupling 
$x=x_t\approx 1/4\pi^2\approx 0.025$ can have an impact within (\ref{lo_const}); the 
bottom quark with $x_b\approx 2\times 10^{-5}$ is dominated by the gauge couplings.
If the two solutions of (\ref{lo_const}) are examined as a function of $x$, the small-$y$ solution becomes negative (and hence non-physical) for $ x>\frac{1}{4}\sqrt{s^2+2rs+3r^2}\approx 0.005 \ll x_t$, indicating that the top-quark Yukawa coupling destabilizes the small Higgs self-coupling solution $y\approx \frac{11}{64}\left(s^2+2rs+3r^2\right)$.  However, (\ref{lo_const}) admits  a   $y\approx 1/11$ solution in the presence of top-quark effects, corresponding to a large Higgs self-coupling.

In the large Higgs self-coupling scenario, $y$ is sufficiently large that higher-order terms in the leading-logarithm expansion of the effective potential can become important.  In particular, the renormalization conditions chosen require terms up to order $L^4$:
\begin{equation}
V_{eff}\equiv\pi^2\phi^4S=\pi ^{2}\phi ^{4}\left( A+BL+CL^{2}+DL^{3}+EL^{4}+\ldots \right)~ ,
\label{eq1}
\end{equation}%
where $A=y+K$ and 
\begin{gather}
V_{eff}^{\prime }(v)=0\Longrightarrow K=-B/2-y~;
\label{vev_cons}
\\
V_{eff}^{(4)}(v)=V_{tree}^{(4)}(v)\Longrightarrow y=\frac{11}{3}B+\frac{35}{3%
}C+20D+16E  ~.\label{eq3}
\end{gather}%
Through the RG equation, the dependence of the
coefficients $\{B,C,D,E\}$  on  the couplings of the theory can be calculated to any given order.
Solution of (\ref{eq3})  then yields the Higgs mass via the expression 
\begin{equation}
m_H^2=V_{eff}^{\prime\prime}(v)= 8\pi^2v^2\left(B+C\right)~.
\label{mass_eq}
\end{equation}
Carrying out this procedure in the absence of the EW coupling effects (but with inclusion of the strong coupling $z$), at leading-logarithm ($LL$) order results in $y=0.0538$ and a viable Higgs mass  of $m_H=216\,{\rm GeV}$ \cite{us1}.  Inclusion of EW corrections has marginal impact on this scenario, resulting in $y=0.0545$ and $m_H=218\,{\rm GeV}$ \cite{us2}.  The value of $y$ is enhanced compared with the conventional symmetry breaking scenario which results in $y=m_H^2/8\pi^2v^2=0.0097$ for a $216\,{\rm GeV}$ Higgs.  Thus the signal of  this radiative symmetry breaking scenario is an order-$220\,{\rm GeV}$ Higgs with an enhanced   Higgs self-coupling.
However,   with a large Higgs-self coupling $y$,  the  perturbative stability of this scenario is not guaranteed. 

\begin{figure}[htb]
\centering
\includegraphics[scale=0.5]{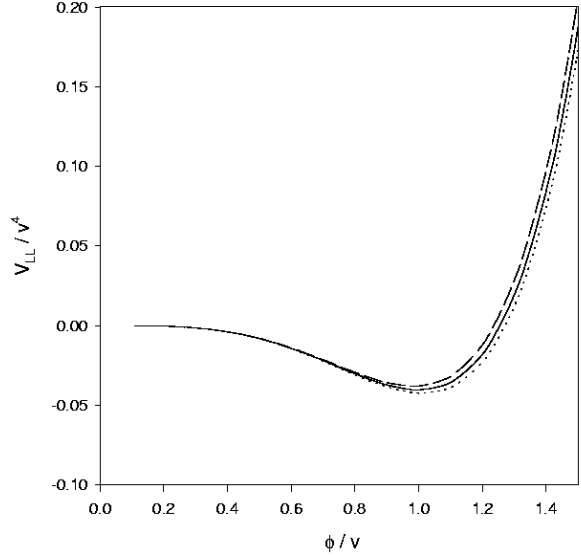}
\caption{
The residual renormalization-scale dependence of the
effective potential.  The  couplants and $\phi(\mu)$ have been RG-evolved from their
values at $\mu = v$ as given in the text.
 The horizontal axis is $\phi(v)/v$, and the vertical axis is
the corresponding $LL$ value of  $V_{eff}/v^4$ for  each curve's choice of $\mu$.
The top, middle and bottom curves respectively
correspond to $\mu = v/2$ ($m_H = 208\, {\rm GeV}$), $\mu = v$ ($m_H = 216\,
{\rm GeV}$), and $\mu = 2v$ ($m_H = 217\, {\rm GeV}$).
}
\label{fig6}
\end{figure}

Eq.\ (\ref{rg_eq}) would appear to eliminate all renormalization scale ($\mu$) dependence from $V_{eff}$.  However, a solution of (\ref{rg_eq}) in the form of a truncated perturbation series will have a small amount of residual $\mu$ dependence.
An estimate of the order-of-magnitude  effect of higher-order perturbative corrections can often be obtained through this residual renormalization scale dependence, since  such residual scale dependence is proportional to the next order corrections and must therefore scale with the coupling of the theory. 
 Varying the renormalization scale in the range $v/2\le \mu \le 2v$  in the $LL$ benchmark SM case
 (including  the strong coupling $z$ and top-quark Yukawa coupling  $x$)  alters $V_{eff}$ as shown in
Figure \ref{fig6}. Numerical determination of $V_{eff}^{\prime\prime}$  at the (shifted) minimum of the effective potentials shown in Figure \ref{fig6} shifts the Higgs mass prediction by approximately $10\,{\rm GeV}$ \cite{us2}.

A  more detailed study of the impact of higher-loop effects in $y$ can be obtained from the scalar field theory projection (SFTP) of the standard model ({\it i.e.} all gauge and Yukawa couplings set to zero).  The SFTP captures the essence of the above $LL$ analyses since it leads to  a solution $y=0.0541$ and $m_H=221\,{\rm GeV}$ very similar to the above scenarios.  In other words, the essential physics is captured by the numerical effects of the Higgs self-coupling. One should note that this does not imply that the underlying theory under examination is purely $\phi^4$ scalar field theory; the SFTP differs by the existence of the underlying EW scale $v=246.2\,{\rm GeV}$ and scalar-gauge boson interactions. The advantage of studying the SFTP is that the RG functions are known to five-loop order via the $\overline{MS}$ results for an $O(4)$-symmetric  massless scalar field theory \cite{rg_ho}: 
\begin{equation}
\beta _{y}=6y^{2}-\frac{39}{2}y^{3}+\frac{10332}{55}%
y^{4}-2698.27y^{5}+47974.7y^{6},\quad \gamma =\frac{3}{8}y^{2}-%
\frac{9}{16}y^{3}+\frac{585}{128}y^{4}-49.8345y^{5} ~. \label{eq7}
\end{equation}%
We consider these RG functions to be compatible with the renormalization conditions 
of Eqs.\ (\ref{vev_cons}) and (\ref{eq3}).
For $y=0.0541$, the perturbative series for these RG functions 
show a slow
monotonic decrease indicative of perturbative stability:
$\beta_y=10^{-3}\left[17.6, -3.09, 1.61, -1.25, 1.20\right]$ and
\\
$\gamma =10^{-3}\left[ 1.10,-0.089,0.039,-0.023\right]$.

The analysis of higher-loop effects in the SFTP  occurs through an iterative approach \cite{us3}.
Consider 
\begin{equation}
V_{eff}=\pi ^{2}\phi ^{4}S,\quad S=y+y\sum_{n=1}^{\infty
}\sum_{m=0}^{n}\;T_{n,m}\;y^{n}{{L}}^{m}~,  \label{eq6}
\end{equation}%
where $K$ is a logarithm-independent counter-term  
\begin{equation}
K=y\sum_{n=1}^\infty y^nT_{n,0}~,~ y\equiv y(v)~.
\end{equation}
 At $LL$ order, the RG equation (\ref{rg_eq}) [up to ${\cal O}\left(y^3\right)$ violations 
 corresponding to the next-order term in $\beta_y$] gives the $LL$ expansion
\begin{equation}
S_{LL}=y+y\sum_{n=1}^\infty \left(yL\right)^n T_{n,n}+K
= \frac{y}{1-3yL} +  K~; \left(T_{n,n}  = 3^n\right)
~,
\label{LL_sol}
\end{equation} 
resulting in
 \begin{equation}
B=3y^2,~C=9y^3,~D=27 y^4,~E=81 y^5~.
\label{LL_coeffs}
\end{equation}
Eq.\ (\ref{LL_sol}) is an exact solution of the RG equation (\ref{rg_eq}) provided $\beta_y$ is
truncated after its lead $6y^2$ term, and provided $\gamma$ is assigned its
corresponding one-loop value of zero [the leading $3y^2/8$ term is two-loop]. The
series values (\ref{LL_coeffs}) lead  via (\ref{eq3}), (\ref{mass_eq}) and (\ref{vev_cons})  to the benchmark $LL$ values $y=0.0541$, $K=-0.05853$ and $m_H=221\,{\rm GeV}$ noted earlier.   

At next-to-leading logarithm ($NLL$) order, the effective potential is expressed as
\begin{equation}
S_{NLL}=\frac{y}{1-3yL} +y^2\sum_{n=1}^\infty \left(yL\right)^{n-1}T_{n,n-1}+\left[K-y^2T_{1,0}\right]
\label{S_NLL}
\end{equation}
where the last term corrects for the inclusion of $T_{1,0}$ in the summation. 
The RG equation (\ref{rg_eq}) [up to ${\cal O}\left(y^4\right)$ violations corresponding to the second
subsequent order in $\beta_y$ and first subsequent order in $\gamma$] can be used to
obtain the following $NLL$ order results for the summation in Eq.\ (\ref{S_NLL}):
\begin{gather}
B=3y^2+\left(6T_{1,0}-\frac{21}{2}\right)y^3~,~
C=9y^3+\left(27T_{1,0}-\frac{621}{8}\right)y^4
\label{NLL_eq1}
\\
D=27 y^4+\left(108T_{1,0}-\frac{801}{2}\right)y^5~,~
E=81 y^5+\left(405T_{1,0}-\frac{28323}{16}\right)y^6~.
\label{NLL_eq2}
\end{gather} 
Eq.~(\ref{vev_cons}) can then be used to express $T_{1,0}$ in terms of $y$ and $K$
\begin{equation}
T_{1,0}=\frac{-4K-6y^2+21y^3-4y}{12y^3}~,
\label{T10_eq}
\end{equation}  
permitting sequential determination $y=0.0538$ via (\ref{eq3}), $T_{1,0}=2.552$ via  (\ref{T10_eq}) and
$m_H=227\,{\rm GeV}$ via (\ref{mass_eq}). Note that the $6\,{\rm GeV}$  shift in the Higgs mass is comparable to the ${\cal O}\left(10\,{\rm GeV}\right)$ residual renormalization scale uncertainty associated with Figure \ref{fig6}. 

Extension to $N^pLL$ order follows the pattern outlined above: the RG equation determines the coefficients $\{B,C,D,E\}$, $T_{n,0}$ is eliminated using  (\ref{vev_cons}), and the numerical value of 
$y$ is found from the constraint (\ref{eq3}) thereby establishing the numerical predictions for $T_{n,0}$ and $m_H$.  The results of this process are shown in Table \ref{res_tab} up to the highest-logarithm  order permitted by the RG functions (\ref{eq7}), and demonstrate remarkable perturbative stability of the Higgs mass and self-coupling \cite{us3}.  
The coefficients $T_{n,0}$ estimated through this iterative scheme have power-law growth characteristic of a perturbative series, providing support for the validity of the iterative approach.

\begin{table}[hbt]
\centering
\begin{tabular}{||c|c|c|c||}
\hline
$n$ & $y\left( v\right) $ & $m_{H}$ & $T_{n,0}$ \\ \hline
$0$ & $0.05414$ & $221.2$ & $1$ 
\\ \hline
$1$ & $0.05381$ & $227.0$ & $2.5521$ 
\\ \hline
$2$ & $0.05392$ & $224.8$ & $-8.1770$ 
\\ \hline
$3$ & $0.05385$ & $226.2$ & $83.211$  
\\ \hline
$4$ & $0.05391$ & $225.0$ & $-1141.8$ \\ \hline
\end{tabular}%
\caption{ Perturbative stability of results inclusive of $N^{n}LL$
contributions from the dominant couplant $y\left( v\right) = \protect%
\lambda \left( v\right) /4\protect\pi ^{2}$ to the SFTP of the SM
effective potential. The quantity $m_{H}$ denotes the VEV-referenced running Higgs boson mass 
 in GeV units 
$\left( v=246.2\,GeV\right) $. }
\label{res_tab}
\end{table}

Further perturbative effects have also been considered \cite{us3}.
The various $N^pLL$ order results  of Table \ref{res_tab} can also be augmented by the $LL$ Yukawa and gauge coupling contributions, resulting in only a marginal shift in the resulting Higgs mass.  Also, the Higgs mass is determined by the full inverse propagator $\Gamma\left(p^2,\mu=v\right)$, which has the simple form 
\begin{equation}
\Gamma\left(p^2,v\right)=\left[1-\left(\frac{3}{4}x(v)-\frac{9}{16}r(v)-
\frac{3}{16}s(v)\right) \log{\left(\frac{p^2}{v^2}\right)} \right]p^2
-V^{\prime\prime}_{eff}(v)~  \label{inv}
\end{equation}
because of the absence of a primitive $\phi^2$ term  in the Lagrangian \cite{inv_prop}.  The corrections from the kinetic term in (\ref{inv}), which follow from a lowest-order RG argument for a massless boson
\cite{politzer},
shift the Higgs mass determined by $V_{eff}^{\prime\prime}(v)$ by less than  $1\,{\rm GeV}$.

Thus we conclude that evidence exists for a perturbatively stable radiatively-generated Higgs mass on the order of $220\mbox{--}230\,{\rm GeV}$, characterized by an approximately five-fold enhancement of the Higgs self-coupling compared with conventional symmetry breaking.  Such an enhanced Higgs self-coupling should be evident via enhancement of specific Higgs processes in the next generation of collider experiments.   

We are grateful for discussions with F.A.\ Chishtie, M.\ Sher, and V.A.\ Miranksy and for support from the Natural Sciences and Engineering Research Council of Canada.

\end{document}